\documentclass[aps,prd,12pt,floatfix,longbibliography,nofootinbib]{revtex4-2}

\usepackage{amsmath,amssymb} % math symbols
\usepackage{bm} % bold math font
\usepackage{graphicx} % for figures
\usepackage{subcaption} % for juxtaposing figures
\usepackage{comment} % allows block comments
\usepackage{textcomp} % This package gives the text quote '

\usepackage{enumitem}
\setlist{noitemsep,leftmargin=*,topsep=0pt,parsep=0pt}

\usepackage{xcolor} % \textcolor{red}{text} will be red for notes
\definecolor{lightgray}{gray}{0.6}
\definecolor{medgray}{gray}{0.4}

\usepackage{hyperref}
\hypersetup{
colorlinks=true,
urlcolor= blue,
citecolor=blue,
linkcolor= blue,
}

\newif\ifptitle
\newif\ifpnumber
\newcounter{para}

\ptitletrue  % comment this line to hide paragraph titles
\pnumbertrue  % comment this line to hide paragraph numbers

\newcommand{\mytitle}{Arbitrarily Negative Energy for Small Kaluza-Klein Bubbles}

\begin{document}

\title{\mytitle}

\author{Gary T. Horowitz}
%\email[]{horowitz@ucsb.edu}
\affiliation{Department of Physics, University of California, Santa Barbara, California 93106, USA}

\author{Guanyu (Ricky) Lu}
%\email[]{ricky@physics.ucsb.edu}
\affiliation{Department of Physics, University of California, Santa Barbara, California 93106, USA}

%\date{\today}

\begin{abstract}
We show that the ADM energy of a Kaluza-Klein bubble of nothing is unbounded from below even if the size of the circle at infinity and the size of the minimal sphere at the bubble are fixed. We demonstrate this by presenting a family of explicit time-symmetric initial data satisfying these boundary conditions with arbitrarily negative energy. In particular, this is true for very small bubbles, which indicates that the standard Kaluza-Klein vacuum is more unstable than previously thought.
\end{abstract}

\maketitle

\baselineskip=16pt

\section{\label{sec:Intro}Introduction}
An important feature of asymptotically flat four-dimensional spacetimes is the positive energy theorem \cite{Schoen:1979rg, Schoen:1979zz, Schoen:1981vd, Witten:1981mf}: if matter satisfies a reasonable local energy inequality, the total ADM energy is non-negative, and $\mathbb{M}^4$, four-dimensional Minkowski spacetime, is the unique solution saturating the bound.

Perhaps the simplest model of unification is Kaluza-Klein theory \cite{Kaluza:1921tu, Klein1926}, which is based on five-dimensional spacetimes that asymptotically approach $\mathbb{M}^4\times S^1$. A natural question is whether there is an analog of the positive energy theorem in this case. If the spacetime admits asymptotically constant spinors, the answer is yes \cite{Zhang:1999ma, Dai_2004, Dai_2005}. However, Witten showed that in general the answer is no, by constructing a nontrivial solution with zero total energy \cite{Witten:1981gj}. Witten's solution has topology $\mathbb{R}^2\times S^2$ on each time slice and represents a ``bubble of nothing" that expands out to infinity. Generalizations of Witten's solution were found with negative energy \cite{Brill:1991qe}, but in these examples, as the energy becomes more negative, the size of the bubble (on a time-symmetric surface) becomes larger.

It is thus natural to ask if there is a bound on the energy of these bubble-of-nothing spacetimes if one fixes the size of the bubble on a time-symmetric surface. Brill and Pfister \cite{Brill:1989di} derived a lower bound to the energy, but their bound was in terms of a radial coordinate in a conformally related flat space which does not have any direct geometric significance. 
 
A more physical and natural choice is to fix the size of the minimal $S^2$ at the bubble (and the $S^1$ at infinity), which is what we consider in this paper. We find that in this case the ADM energy is not bounded from below by constructing an explicit solution to the time-symmetric initial value constraint with parameters that can be varied to achieve an arbitrarily negative energy. We further show that our result is in agreement with the result of Brill and Pfister \cite{Brill:1989di}, since their bound also becomes arbitrarily negative as $ E \rightarrow -\infty$.

Since the positive energy theorem does not hold for Kaluza-Klein theory (unless one requires asymptotically constant spinors), the standard vacuum, $\mathbb{M}^4\times S^1$, is nonperturbatively unstable to decay to another zero energy state. In fact, Witten \cite{Witten:1981gj} constructed an instanton which describes this decay for his solution. The rate of decay is related to the Euclidean action of this instanton, and its scale is set by the minimal size of the bubble. Now that we have small bubbles with negative energy, we can certainly construct many more solutions with small bubbles and zero energy. Although we do not have the corresponding instantons, it suggests that the Kaluza-Klein vacuum might decay much faster than previously expected.

The fact that small Kaluza-Klein bubbles can have arbitrarily negative energy raises a potential problem for the AdS/CFT correspondence.  Since anti-de Sitter (AdS) spacetime is approximately flat on scales much smaller than the AdS radius, it would appear that one could insert a small bubble of nothing with very negative energy and  make the total energy arbitrarily negative. This would contradict the fact that the dual CFT has a Hamiltonian that is bounded from below. Fortunately, we show that this does not happen. Analogous initial data with a negative cosmological constant has a total energy which does not become arbitrarily negative, in agreement with the AdS/CFT correspondence. One reason for the difference is that in the asymptotically flat case, the $S^1$ approaches a fixed size and the energy is obtained from the $1/r$ correction to the metric. In contrast, in a locally AdS spacetime with conformal boundary $S^1\times S^2\times\mathbb{R}$, the $S^1$ becomes large and the energy is obtained from the $1/r^2$ correction to the metric.

 In the next section we introduce the form of the initial data for Kaluza-Klein theory that we will consider. In section \ref{sec:Gen} we derive a general formula for the ADM energy (with some details relegated to Appendix \ref{app:MassGen}). Section \ref{sec:Sol} gives our specific family of initial data and shows that it has unbounded energy for fixed bubble size. The following section shows that this satisfies some consistency checks including the Brill-Pfister bound. Section \ref{sec:AdS} shows that analogous initial data for asymptotically AdS spacetimes has bounded energy, and the final section has some concluding remarks. Appendix \ref{app:PolA} presents another family of Kaluza-Klein initial data with similar properties, but smaller curvature at the bubble.

\section{\label{sec:Setup}Setup}
We will set the four-dimensional Newton's constant to 1. The ADM energy of a five-dimensional spacetime that asymptotically approaches $\mathbb{M}^4\times S^1$ is given by \cite{Bombelli:1986sb,Deser:1988fc} 
\begin{equation}\label{eq:ADM}
    E_{\text{ADM}}=\frac{1}{16\pi\times2\pi R}\int\left(\nabla^jh_{ij}-\nabla_ih^{j}_{j}\right)dS^i,
\end{equation}
where $R$ is the radius of $S^1$ at infinity, $h_{ij}=g_{ij}-\eta_{ij}$ is the deviation of the metric from the flat metric on a time slice, the covariant derivatives are taken with respect to $\eta_{ij}$, the flat metric on $\mathbb{R}^3\times S^1$, and $S^i$ is the outgoing unit normal vector on $S^2\times S^1$ at infinity. As usual, the ADM energy does not depend on the choice of the time slice. Essentially the only difference between Eq.\ \eqref{eq:ADM} and the usual one in four dimensions is the factor of $2\pi R$ in the denominator which arises since the
five-dimensional Newton's constant is $G_5 = 2\pi R G_4 = 2\pi R$. More generally, for a spacetime that asymptotically approaches $\mathbb{M}^4\times K$ for an internal manifold $K$, this factor is the volume of $K$. From now on, we will omit the subscript on the energy.

We consider time-symmetric initial data with $SO(3)\times U(1)$ symmetry. 
The metric can be written as
\begin{equation}\label{eq:metric}
    ds^2=f_1(\rho)d\rho^2+\rho^2d\Omega^2+f_2(\rho)d\phi^2,
\end{equation}
where $d\Omega^2$ is the round metric on the unit $S^2$, and for large $\rho$,
\begin{equation}\label{eq:f12}
    f_1(\rho)=1+\frac{2m}{\rho}+\mathcal{O}\left(\rho^{-2}\right),\quad f_2(\rho)=1+\frac{4\mu}{\rho}+\mathcal{O}\left(\rho^{-2}\right).
\end{equation}
Using Eq.\ \eqref{eq:ADM}, one can calculate the ADM energy of this metric to be
\begin{equation}\label{eq:mass}
    E=m+\mu.
\end{equation}

Now we focus on the Kaluza-Klein bubble spacetimes. These are spacetimes in which the $S^1$ pinches off so that $f_2(\rho_0)=0$ for some $\rho_0>0$, the radius of the minimal $S^2$. The radial coordinate $\rho$ takes values in $[\rho_0,\infty)$ and the functions $f_1(\rho)$ and $f_2(\rho)$ are smooth positive functions of $\rho$ for $\rho\in(\rho_0,\infty)$. Near $\rho=\rho_0$, to leading order in $\rho-\rho_0$, 
\begin{equation}\label{eq:f12ab}
    f_1(\rho)=a(\rho-\rho_0)^{-1},\quad f_2(\rho)=b(\rho-\rho_0),
\end{equation}
for some $a,b\in\mathbb{R}$. 
%In order for the metric to be free of curvature singularities at $\rho=\rho_0$, we need $p=-1$, $q=1$. 
For periodic $\phi$ with periodicity $2\pi R$, in order for the metric to be free of conical singularities at $\rho=\rho_0$,
\begin{equation}\label{eq:sc}
    \frac{bR^2}{4a}=1.
\end{equation}

\section{\label{sec:Gen}General Expression for the Energy}
The goal now is to find  functions $f_1(\rho)$ and $f_2(\rho)$, satisfying the boundary conditions and the Hamiltonian constraint, with some parameters that can be varied to obtain an arbitrarily negative ADM energy. The extrinsic curvature of the time-symmetric slice vanishes, so the Hamiltonian constraint reduces to the constraint that the four-dimensional Ricci scalar of the metric, Eq.\ \eqref{eq:metric}, vanishes. The Ricci scalar involves $f_2''(\rho)$ but only $f'_1(\rho)$, so it is convenient to specify some $f_2(\rho)$ with some parameters and solve for $f_1(\rho)$.

For simplicity of calculations, we rewrite the metric functions as
\begin{equation}\label{eq:mab}
    f_1(\rho)=\left(\alpha(\rho)\beta(\rho)\right)^{-1},\quad f_2(\rho)=\alpha(\rho).
\end{equation}
Define the following functions
\begin{subequations}\label{eq:h12}
\begin{align}
    h_1(\rho)&=\alpha(\rho)+2\rho\alpha'(\rho)+\frac{\rho^2\alpha''(\rho)}{2},\\
    h_2(\rho)&=\rho\left(\alpha(\rho)+\frac{\rho\alpha'(\rho)}{4}\right).
\end{align}
\end{subequations}
The four-dimensional Ricci scalar is
\begin{equation}
    -\frac{\rho^2}{2}{}^{(4)}R=-1+h_1(\rho)\beta(\rho)+h_2(\rho)\beta'(\rho).
\end{equation}
The general solution to the Hamiltonian constraint for $\beta(\rho)$ is then given by
\begin{equation}\label{eq:gensol}
    \beta(\rho)=e^{-\int_{\rho_0}^{\rho}\frac{h_1(x)}{h_2(x)}dx}\left(c+\int_{\rho_0}^{\rho}\frac{e^{\int_{\rho_0}^{y}\frac{h_1(x)}{h_2(x)}dx}}{h_2(y)}dy\right),
\end{equation}
where $c$ is the integration constant determined by the smoothness constraint, Eq.\ \eqref{eq:sc}, to be
\begin{equation}\label{eq:c}
    c=\frac{4}{R^2\left(\alpha'(\rho_0)\right)^2}.
\end{equation}
For large $\rho$, we can expand $\alpha(\rho)$ as
\begin{equation}\label{eq:ae}
    \alpha(\rho)=1+\sum_{i=1}^{\infty}\alpha_i\rho^{-i}.
\end{equation}
The ADM energy is then derived in appendix \ref{app:MassGen} to be
\begin{subequations}\label{eq:massgen}
\begin{align}
    E&=\frac{\alpha_1}{8}+\frac{\rho_0}{2}-\frac{2\rho_0}{R^2e^{I_1}\left(\alpha'(\rho_0)\right)^2}-\frac{I_2}{2},\\
    I_1&=\int_{\rho_0}^{\infty}\left(\frac{h_1(x)}{h_2(x)}-\frac{1}{x}\right)dx,\\
    I_2&=\int_{\rho_0}^{\infty}\left(\frac{\rho_0e^{\int_{\rho_0}^{y}\frac{h_1(x)}{h_2(x)}dx}}{e^{I_1}h_2(y)}-1\right)dy.
\end{align}
\end{subequations}

Because of the appearance of $h_2(\rho)$ in the denominators, we require that $h_2(\rho)>0$ $\forall\rho\geq\rho_0$. It is clear that $h_2(\rho)>0$ at and near $\rho_0$ and towards infinity.  While in general $h_2(\rho)$ could vanish somewhere in the middle, this does not happen for our choice of $\alpha(\rho)$ to be presented in the next section.

\section{\label{sec:Sol}Arbitrarily Negative ADM Energy}
In this section, we present a function $\alpha(\rho)$ with some parameters that can be varied to make the ADM energy, Eq.\ \eqref{eq:massgen}, arbitrarily negative with the size of the bubble $\rho_0$ fixed.

Consider the function
\begin{equation}
    \alpha(\rho)=\frac{\rho-\rho_0}{\rho-\rho_0+d}\exp\left(-\frac{p}{\rho}\right),
\end{equation}
where $\rho\geq\rho_0>0$, $d>0$, $p\in\mathbb{R}$, which clearly satisfies our desired boundary conditions. We also need to ensure that $h_2(\rho)>0$ $\forall\rho\geq\rho_0$. For this function,
\begin{equation}
    h_2(\rho)=\left(4+\frac{\rho}{\rho-\rho_0}-\frac{\rho}{\rho-\rho_0+d}+\frac{p}{\rho}\right)\frac{\rho\alpha(\rho)}{4}.
\end{equation}
Clearly, for $p\geq0$, the regime we will be interested in, $h_2(\rho)>0$ $\forall\rho\geq\rho_0$, as required. 

Now let us analyze the ADM energy of this metric. We can easily calculate
\begin{equation}\label{eq:ald}
    \alpha'(\rho_0)=\frac{e^{-p/\rho_0}}{d},\quad\alpha_1=-p-d,\quad\alpha_2=\frac{(p+d)^2+d^2-2d\rho_0}{2},
\end{equation}
where $\alpha_1$ and $\alpha_2$ are the coefficients of the asymptotic expansion of $\alpha(\rho)$ as in Eq.\ \eqref{eq:ae}. To study the integrals, we use the partial fraction decomposition to write
\begin{equation}\label{eq:PFD}
    \frac{h_1(\rho)}{h_2(\rho)}=\frac{2p}{\rho^2}-\frac{4}{\rho}-\frac{4}{\rho-\rho_0+d}+\sum_{i=1}^{3}\frac{p_i}{\rho-\rho_0-\rho_i},
\end{equation}
where $\rho_i$ are the roots of the polynomial
\begin{equation}
    x^3+\frac{5d+4\rho_0+p}{4}x^2+\frac{6\rho_0d+dp}{4}x+\frac{\rho_0^2d}{4},\label{eq:pol}
\end{equation}
and $p_i$ are the following functions of the respective $\rho_i$
\begin{equation}
    p_i=\frac{36\rho_i^2+(32d+16\rho_0+4p)\rho_i+12d\rho_0+2dp}{12\rho_i^2+(10d+8\rho_0+2p)\rho_i+6d\rho_0+dp}\label{eq:pow}.
\end{equation}
We can then calculate the integral
\begin{equation}
    e^{\int_{\rho_0}^{\rho}\frac{h_1(x)}{h_2(x)}dx}=\left(\frac{\rho_0d}{\rho(\rho-\rho_0+d)}\right)^4\prod_{i=1}^{3}\left(\frac{\rho-\rho_0-\rho_i}{-\rho_i}\right)^{p_i}e^{\frac{2p}{\rho_0}-\frac{2p}{\rho}}.
\end{equation}
It is clear from Eq.\ \eqref{eq:h12} that $h_1/h_2$ goes like $1/\rho$ asymptotically. Hence, Eq.\ \eqref{eq:PFD} implies
\begin{equation}
    \sum_{i=1}^{3}p_i=9.
\end{equation}
Hence,
\begin{equation}
    e^{I_1}=\rho_0^5d^4\prod_{i=1}^{3}\left(-\rho_i\right)^{-p_i}e^{\frac{2p}{\rho_0}}.
\end{equation}

%Now we want to consider the scenario of large $p$, $p\gg0$. From the polynomial, Eq.\ \eqref{eq:pol}, by Vieta's formulas, we see that for large $p$, to leading order in $p$, one root, which we label as $\rho_1$ without loss of generality, is $\mathcal{O}(p)$ while the other two are $\mathcal{O}(1)$, which we label $\rho_2$ and $\rho_3$. Further, using these growth rates, we can expand the polynomial to obtain that
Now we want to consider the scenario of large $p$, $p\gg0$. Using Vieta's formulas, we can expand the polynomial, Eq.\ \eqref{eq:pol}, to obtain that
\begin{subequations}
\begin{align}
    \rho_1&=-\frac{p}{4}-\rho_0-\frac{d}{4}+\frac{d(2\rho_0-d)}{p}+\mathcal{O}\left(p^{-2}\right),\\
    \rho_2&=-d+\frac{(\rho_0-d)^2}{p}+\mathcal{O}\left(p^{-2}\right),\\
    \rho_3&=-\frac{\rho_0^2}{p}+\mathcal{O}\left(p^{-2}\right).
\end{align}
\end{subequations}
Therefore,
\begin{equation}
    p_1=5+\mathcal{O}\left(p^{-1}\right),\quad p_2=2+\mathcal{O}\left(p^{-1}\right),\quad p_3=2+\mathcal{O}\left(p^{-1}\right).
\end{equation}
Thus, we obtain
\begin{equation}\label{eq:eI1}
    e^{I_1}=2^{10}\rho_0d^2p^{-3}e^{\frac{2p}{\rho_0}}\left[1+\mathcal{O}\left(p^{-1}\right)\right].
\end{equation}
Hence, its contribution to the energy, Eq.\ \eqref{eq:massgen}, is
\begin{equation}\label{eq:dominantE}
    -\frac{2\rho_0}{R^2e^{I_1}\left(\alpha'(\rho_0)\right)^2}=-\frac{p^3}{2^{9}R^{2}}\left[1+\mathcal{O}\left(p^{-1}\right)\right].
\end{equation}
Note that it is negative and grows like $p^3$.

\begin{figure}[h]
    \includegraphics[clip=true,width=0.5\columnwidth]{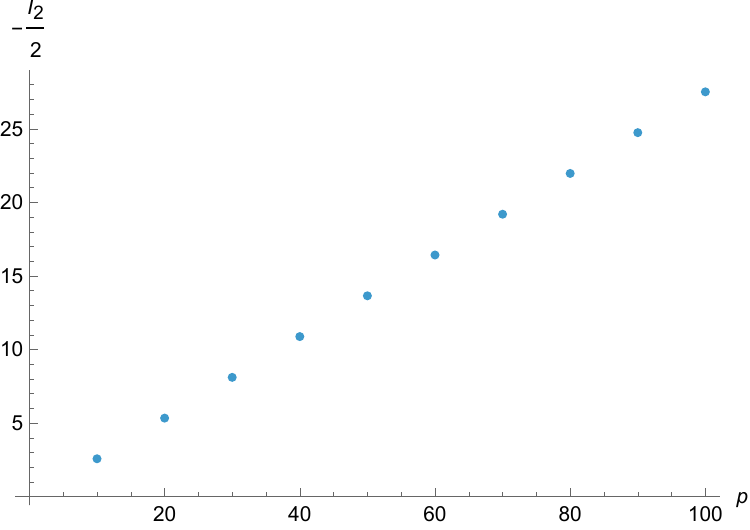}
    \caption{A plot of $-\frac{I_2}{2}$ vs.\ $p$, with $\rho_0=d=1$.}
    \label{fig:I2}
\end{figure}

The contribution to the energy from the $I_2$ term in Eq.\ \eqref{eq:massgen} can be analyzed without further explicit integrations.  From Eqs.\ \eqref{eq:h12} and \eqref{eq:eI1}, we see that the first term of the integrand of $I_2$ is exponentially suppressed near $\rho_0$. Since this first term is positive, the most negative $I_2$ can be is when this term stays close to zero until large $\rho$ when it approaches one as shown in Eq.\ \eqref{eq:I2i}.  Since it must approach one by $\rho\sim\sqrt{|\alpha_2|}$, and for large $p$, $\sqrt{|\alpha_2|}\sim p$ from Eq.\ \eqref{eq:ald}, we see that $I_2$ is at most $\mathcal{O}(p)$ for large $p$. A numerical calculation\footnote{We used $1/\rho$ as the integration variable. Because the integrand is the difference of two divergent terms and is obtained through numerical integration, a cutoff needs to be imposed for small $1/\rho$ for the outer integral (unless the inner one has infinite precision) and the (negligible) asymptotic contribution can be obtained from the asymptotic behavior, Eq.\ \eqref{eq:I2i}.} of $I_2$ is shown in Fig.\ \ref{fig:I2} where we see that $I_2$ is indeed linear in $p$.

Moreover, from Eq.\ \eqref{eq:ald}, we see that the contribution of the $\alpha_1$ term to the energy is also of $\mathcal{O}(p)$. Hence, to leading order in $p$ for large $p$, the energy is given by \eqref{eq:dominantE}
\begin{equation}
    E=-\frac{p^3}{2^{9}R^{2}}+\mathcal{O}\left(p^{2}\right).
\end{equation}
Thus, as $p$ becomes larger, the ADM energy of the metric becomes more negative. In other words, the ADM energy is unbounded from below for any choice of $\rho_0$ and $R$.

\section{\label{sec:Consistency}Consistency Checks}
\subsection{\label{subsec:BP}Brill-Pfister Bound}
We can rewrite the metric, Eq.\ \eqref{eq:metric}, via a coordinate transformation $r=r(\rho)$ as
\begin{equation}
    ds^2=\psi^4(r)(dr^2+r^2d\Omega^2)+V^2(r)d\phi^2,
\end{equation}
which is related to Eq.\ \eqref{eq:metric} by
\begin{equation}\label{eq:coord}
    \sqrt{f_1(\rho)}d\rho=\psi^2(r)dr,\quad\rho=\psi^2(r)r,\quad f_2(\rho)=V^2(r).
\end{equation}
Let $B=r(\rho_0)$. Brill and Pfister showed that \cite{Brill:1989di}
\begin{equation}\label{eq:BPbound}
    E>-2B.
\end{equation}
In other words, the total energy is always bounded from below by the radial coordinate in a conformally related flat metric. Since our initial data has arbitrarily negative $E$ (for fixed $\rho_0$)  it better be that $B$ can be arbitrarily positive. 

\begin{figure}[h]
    \includegraphics[clip=true,width=0.7\columnwidth]{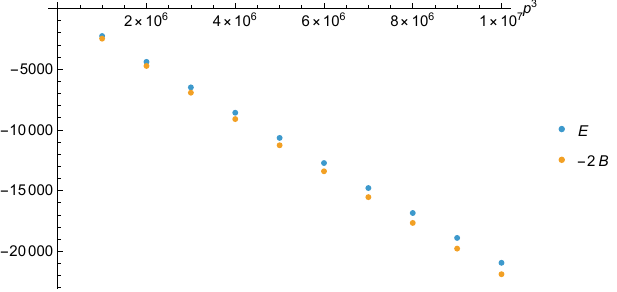}
    \caption{A plot of $E$ and $-2B$ vs.\ $p^3$, with $\rho_0=R=d=1$.}
    \label{fig:EB}
\end{figure}

To compute $B$ for our initial data,  note that Eq.~\eqref{eq:coord} implies 
$\sqrt{f_1} d\rho/\rho = dr/r$, which is easily integrated to yield 
\begin{equation}\label{eq:B1}
    B=r\exp\left(-\int_{\rho_0}^{\rho}\frac{\sqrt{f_1(x)}}{x}dx\right).
\end{equation}
Since our asymptotic boundary conditions require $r=\rho$ (to leading order) for large $\rho$,
\begin{equation}\label{eq:B}
    B=\rho_0\exp\left(\int_{\rho_0}^{\infty}\frac{1-\sqrt{f_1(x)}}{x}dx\right).
\end{equation}
Using this, we have computed $B$ and checked that Eq.\ \eqref{eq:BPbound} is indeed satisfied. As shown in Fig.\ \ref{fig:EB}, both $E$ and $B$ are $\mathcal{O}\left(p^3\right)$ and $E>-2B$.

The following argument gives a rough, intuitive explanation for why $B$ grows as $E$ becomes more negative. In order for $B$ to become large, we need $f_1(\rho)$ to stay close to $0$ for a large range before going back to $1$. From Eqs.~\eqref{eq:f12ab}-\eqref{eq:mab} and \eqref{eq:ald}, our initial data with large negative energy (large $p$) has very small $b = \alpha'(\rho_0)$ and hence small $a$. So $f_1(\rho)$ stays close to $0$ for $\rho$ near $\rho_0$. Asymptotically, we know $f_1 = 1+2m/\rho +\cdots$ where $m = \mathcal{O}\left(p^3\right)$. If we assume that $f_1$ stays small until $\rho\sim|m|$, then $B=\mathcal{O}(m)= \mathcal{O}\left(p^3\right)$.

\subsection{\label{subsec:Acceleration}Initial Acceleration of the Bubble}
Since our initial data is time symmetric, the initial velocity of the bubble is zero. However, the initial acceleration is nonzero and given by \cite{Corley:1994mc}
\begin{equation}\label{eq:A}
    \ddot{A}=-8\pi{}^{(4)}R_{\theta\theta}(\rho_0),
\end{equation}
where ${}^{(4)}R_{\theta\theta}(\rho_0)$ is the relevant component of the Ricci tensor from the four-dimensional metric, Eq.\ \eqref{eq:metric}, evaluated at $\rho=\rho_0$, and we assume that the full five-dimensional theory has a vanishing stress-energy tensor. If bubbles with negative energy started to collapse, they could not form black holes (which have positive energy). Instead, they might form naked singularities and violate the weak cosmic censorship hypothesis. It has been shown \cite{Corley:1994mc} that all the negative energy Brill-Horowitz bubbles \cite{Brill:1991qe} start out expanding.

We can calculate
\begin{equation}\label{eq:R4}
    {}^{(4)}R_{\theta\theta}(\rho_0)=1-\rho_0\alpha'(\rho_0)\beta(\rho_0)=1-\frac{4\rho_0}{R^2\alpha'(\rho_0)},
\end{equation}
where we have used Eq.\ \eqref{eq:c}. Hence, for our solution,
\begin{equation}
    \ddot{A}=-8\pi\left(1-\frac{4\rho_0d}{R^2}e^{\frac{p}{\rho_0}}\right).
\end{equation}
Thus, our Kaluza-Klein bubbles with very negative ADM energy (large $p$) start out expanding at the moment of time symmetry.

In fact, from the analysis in the previous subsection, (at least very) negative energy implies $\alpha'(\rho_0)$ is small. Therefore, from Eqs.\ \eqref{eq:A} and \eqref{eq:R4}, it is not surprising that for spacetimes with very negative energy, the bubble starts out expanding at the moment of time symmetry.
It remains an open question whether any time-symmetric Kaluza-Klein bubble spacetime with a negative energy always starts out expanding.

\section{\label{sec:AdS}Contrast with A\MakeLowercase{d}S}
In view of the AdS/CFT correspondence, one might wonder whether a similar result holds if a negative cosmological constant is introduced. Since the Kaluza-Klein bubble has topology $S^2\times S^1\times\mathbb{R}$ asymptotically, the analog would be spacetimes that are locally asymptotically AdS  with a conformal boundary that has topology $S^2\times S^1\times\mathbb{R}$ instead of the standard $S^3\times\mathbb{R}$. Static solutions with these boundary conditions were found in \cite{Copsey:2006br}, where it was shown that there are two classes of static solutions: there always exist solutions where the $S^1$ does not pinch off, and if the radius of the $S^1$ is small enough compared to the $S^2$ there are two additional solutions where the $S^1$ does pinch off. It was conjectured that the static solution with the lowest energy minimizes the energy among all solutions with these boundary conditions. If the analog of our Kaluza-Klein initial data had arbitrarily negative energy, it would clearly violate this conjecture (and a more refined version in \cite{Horowitz:2024hqq}) and cause a serious problem for the AdS/CFT correspondence. We show that this is not the case.

Let us again consider spherically symmetric and time-symmetric spacetimes, with the metric on the time-symmetric slice written as in Eqs.\ \eqref{eq:metric} and \eqref{eq:mab}. For a bubble of size $\rho_0>0$, the discussion in Sec.\ \ref{sec:Setup} after Eq.\ \eqref{eq:mass}, in particular the smoothness constraint, Eq.\ \eqref{eq:sc}, still applies. However, the asymptotics are different. Let $\ell>0$ be the $\mathrm{AdS}$ radius, related to the cosmological constant $\Lambda$ via $\Lambda=-6/\ell^2$. For $\rho\gg\ell$, the metric functions given in Eq.\ \eqref{eq:mab} have the following asymptotics \cite{Copsey:2006br, Horowitz:2024hqq}:
\begin{subequations}\label{eq:ab}
\begin{align}
    \frac{\ell^2\alpha(\rho)}{\rho^2}&=1+\frac{\ell^2}{2\rho^2}+\frac{\ln\left(\rho/\ell\right)\ell^4}{12\rho^4}+\frac{A_4\ell^4}{\rho^4}+\mathcal{O}\left(\frac{\ln\left(\rho/\ell\right)\ell^5}{\rho^{5}}\right),\\
    \beta(\rho)&=1+\frac{\ell^2}{6\rho^2}+\frac{\ln\left(\rho/\ell\right)\ell^4}{12\rho^4}+\frac{B_4\ell^4}{\rho^4}+\mathcal{O}\left(\frac{\ln\left(\rho/\ell\right)\ell^5}{\rho^{5}}\right).
\end{align}
\end{subequations}
This is the same as the static solutions and is required for consistency with the static boundary metric. The energy can be calculated with the usual counterterm subtraction \cite{Balasubramanian:1999re} (or background subtraction \cite{Hawking:1995fd}) method to be \cite{Copsey:2006br, Horowitz:2024hqq}
\begin{equation}
    E=\frac{\pi R\ell}{2G_5}\left(A_4-3B_4+C_0\right),
\end{equation}
where $2\pi R$ is the period of $S^1$ as before, $G_5$ is the five-dimensional Newton's constant, and $C_0$ is a constant coming from the counterterm subtraction (or the choice of the background).

The Hamiltonian constraint ${}^{(4)}R=2\Lambda$, where ${}^{(4)}R$ is the Ricci scalar of the time-symmetric slice, is given by
\begin{equation}
    h_2(\rho)\beta'(\rho)+h_1(\rho)\beta(\rho)=1+\frac{6\rho^2}{\ell^2},
\end{equation}
where $h_1(\rho)$ and $h_2(\rho)$ are given in Eq.\ \eqref{eq:h12}. The general solution for $\beta(\rho)$ for a given $\alpha(\rho)$ is
\begin{equation}
    \beta(\rho)=e^{-\int_{\rho_0}^{\rho}\frac{h_1(x)}{h_2(x)}dx}\left[c+\int_{\rho_0}^{\rho}\frac{e^{\int_{\rho_0}^{y}\frac{h_1(x)}{h_2(x)}dx}}{h_2(y)}\left(1+\frac{6y^2}{\ell^2}\right)dy\right],
\end{equation}
where $c$ is the integration constant given in Eq.\ \eqref{eq:c}. The energy can be derived in a way analogous to the derivation in appendix \ref{app:MassGen} to be
\begin{subequations}
\begin{align}
    E&=\frac{\pi R\ell}{2G_5}\left[2A_4-\frac{3\rho_0^4}{\ell^4}\left(\frac{4}{R^2e^{I_1}\left(\alpha'(\rho_0)\right)^2}+I_2\right)+\frac{3\rho_0^4}{\ell^4}+\frac{2\rho_0^2}{\ell^2}+\frac{\ln(\rho_0/\ell)}{3}+\frac{11}{24}+C_0\right],\\
    I_1&=\int_{\rho_0}^{\infty}\left(\frac{h_1(x)}{h_2(x)}-\frac{4}{x}\right)dx,\\
    I_2&=\int_{\rho_0}^{\infty}\left[\frac{\exp\left(\int_{\rho_0}^{y}\frac{h_1(x)}{h_2(x)}dx\right)}{e^{I_1}h_2(y)}\left(1+\frac{6y^2}{\ell^2}\right)-\frac{4y^3}{\rho_0^4}\left(1+\frac{\ell^2}{3y^2}+\frac{\ell^4}{36y^4}\right)\right]dy.
\end{align}
\end{subequations}

\begin{figure}[h]
\begin{subfigure}{0.45\textwidth}
\includegraphics[width=\linewidth]{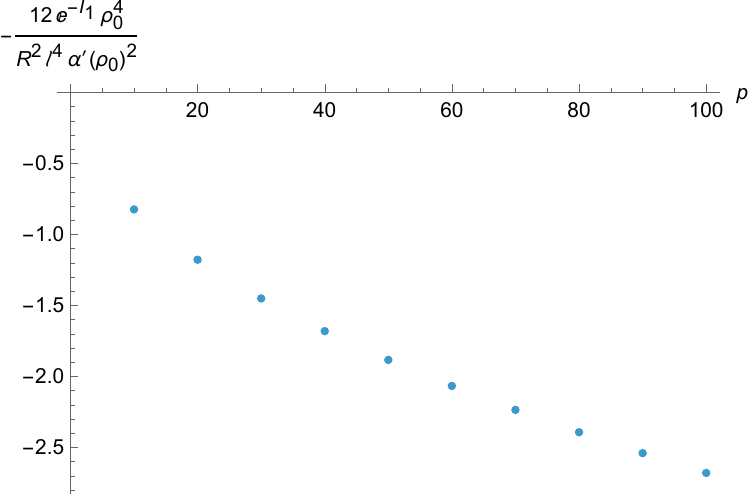} 
\caption{The $I_1$ contribution to $E$ vs.\ $p$}
%$-12R^{-2}(\rho_0/\ell)^4e^{-I_1}\left(\alpha'(\rho_0)\right)^{-2}$ vs.\ $p$}
\label{fig:subim1}
\end{subfigure}
\qquad
\begin{subfigure}{0.45\textwidth}
\includegraphics[width=\linewidth]{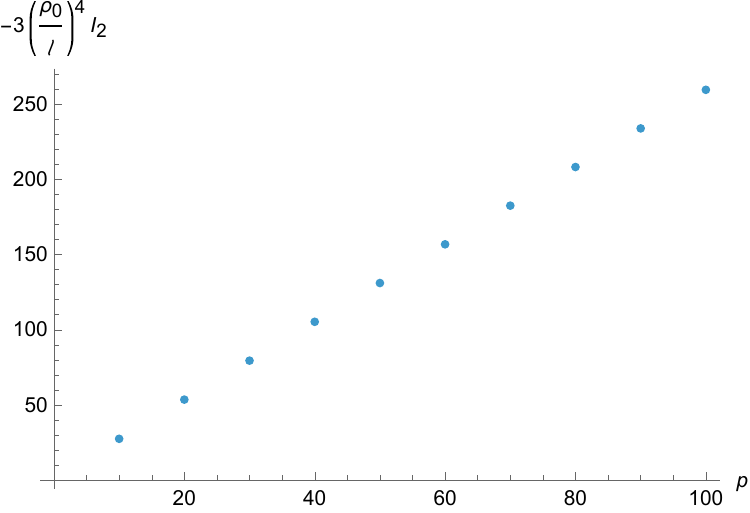}
\caption{The $I_2$ contribution to $E$ vs.\ $p$}
%$-3(\rho_0/\ell)^4I_2$ vs.\ $p$}
\label{fig:subim2}
\end{subfigure}
\caption{Plots of the integral contributions to the energy, with $R=\ell=d=1$, $\rho_0=1/4$. }
\label{fig:MI_AdS}
\end{figure}

Now we would like to see whether the energy can be arbitrarily negative with a negative cosmological constant. Consider the following function, which satisfies Eq.\ \eqref{eq:ab} and is analogous to the function in the previous case with a vanishing cosmological constant:
\begin{equation}
    \alpha(\rho)=\exp\left(-\frac{p\ell^4}{\rho^4}\right)\frac{((\rho/\ell)^2-(\rho_0/\ell)^2)^2+d((\rho/\ell)^2-(\rho_0/\ell)^2)+\ln(\rho/\rho_0)/12}{(\rho/\ell)^2-(\rho_0/\ell)^2+d-(\rho_0/\ell)^2-1/2},
\end{equation}
where $\rho\geq\rho_0>0$, $d>(\rho_0/\ell)^2+1/2$, $p\in\mathbb{R}$. Like before, we need to ensure that $h_2(\rho)>0$ $\forall\rho\geq\rho_0$. While the expression for $h_2(\rho)$ now is very complicated, it is clear that large and positive $p$ is sufficient, which is the analogous regime of interest. We can calculate
\begin{subequations}
\begin{align}
    A_4&=\frac{\rho_0^4}{\ell^4}+\frac{(1-d)\rho_0^2}{\ell^2}+\frac{1}{4}-\frac{\ln\left(\rho_0/\ell\right)}{12}-p-\frac{d}{2},\\
    \alpha'(\rho_0)&=\frac{\left(24d\rho_0/\ell+\ell/\rho_0\right)e^{-p\ell^4/\rho_0^4}}{6l(2d-2(\rho_0/\ell)^2-1)}.
\end{align}
\end{subequations}
Note that for large $p$, $A_4\approx-p$. This time, the integrals are too complicated to evaluate analytically, so we use numerical integration. As shown in Fig.\ \ref{fig:MI_AdS}, the contribution from the term involving $e^{-I_1}\left(\alpha'(\rho_0)\right)^{-2}$ to $E$ (which is negative) is less than linear in $p$ for large $p$, while the contribution of the term involving $I_2$ to $E$ (which is positive) is linear in $p$ for large $p$, having an absolute value larger than that of the term involving $A_4$.

\begin{figure}[h]
    \includegraphics[clip=true,width=0.5\columnwidth]{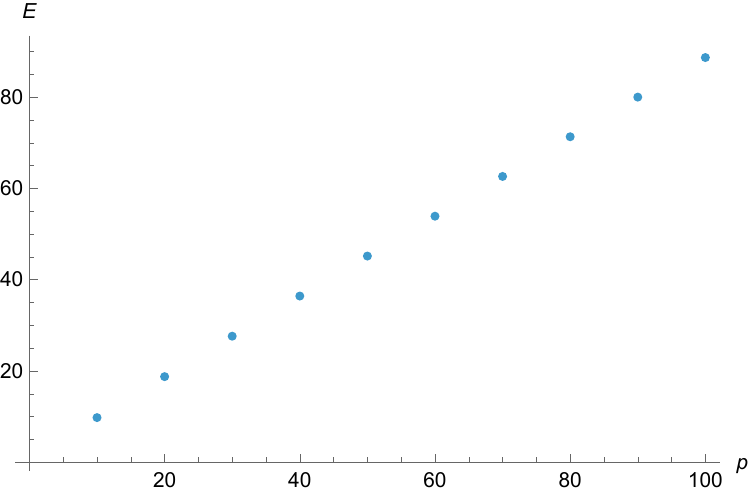}
    \caption{A plot of $E$ vs.\ $p$, with $R=\ell=G_5=d=1$, $\rho_0=1/4$, $C_0=-7/24$. (The linear behavior is modified at small $p$.)}
    \label{fig:Ep}
\end{figure}

Since the other contributions to $E$ are independent of $p$, the energy does not become negative when $p$ becomes large. This is clearly shown in Fig.\ \ref{fig:Ep} for the parameters $R=\ell=G_5=d=1$, $\rho_0=1/4$. (We have also set the constant $C_0$ in the energy to $-7/24$ which is the value for the background subtraction method \cite{Copsey:2006br}.)  We tried other values of $\rho_0/\ell$, changing the size of the bubble, and the term involving $I_1$ is always subdominant while the slope of $(\rho_0/\ell)^4I_2$ for large $p$ does not seem to depend on $\rho_0/\ell$ so it always dominates the term involving $A_4$. Thus, unlike the previous case with a vanishing cosmological constant, the large $p$ limit of this family of metrics does not produce arbitrarily negative energy.

We have tried other scenarios and other functions, but none of them produce arbitrarily negative energy. While we do not have a proof, we are led to believe that the energy is bounded from below, as one would expect from AdS/CFT. The metric functions have different asymptotics and the energy comes from different terms in the two cases, so it comes as no surprise that the energy can be unbounded in one case and bounded in the other. It should also be noted that our initial data with $E \ll 0$ and small bubble size $\rho_0$ in Kaluza-Klein theory differs from the standard Kaluza-Klein vacuum over a large region whose scale is not set by $\rho_0$ but rather by $p$ which is related to the energy. So there is no reason to expect the AdS analog to have a similar behavior.

\section{\label{sec:Conclusion}Conclusion}
We have presented an explicit family of time-symmetric initial data for five-dimensional Kaluza-Klein theory describing bubbles of nothing with negative ADM energy. Unlike previous examples, the energy is unbounded from below even when the size of the bubble, i.e., the size of the minimal $S^2$ (and the size of the $S^1$ at infinity) are fixed. Since this is true for arbitrarily small bubbles, it suggests that the standard Kaluza-Klein vacuum is more unstable than previously thought. 

To obtain further evidence for this instability, the first step is to construct zero energy initial data that resembles the Kaluza-Klein vacuum outside a small region. This is because quantum tunneling events conserve energy (so we need $E=0$) and the rate is determined by the action of an instanton which should be small if the metric is only changed in a small region. If one considers a theory with fermions and imposes boundary conditions that the fermions be periodic around the $S^1$ at infinity, then there is no instability. This is because fermions on a bubble of nothing spacetime are always antiperiodic around this $S^1$ \cite{Witten:1981gj}.

In our example, as $E\rightarrow-\infty$, the curvature at the bubble grows exponentially in $|E|^{1/3} $. This is because $E \propto -p^3$ but the metric is exponential in $p$.  There are other more complicated examples showing that the energy is unbounded from below where the curvature only grows polynomially with the energy, as explained in appendix \ref{app:PolA}. Although the energy can be arbitrarily negative classically, once the curvature exceeds the Planck scale we expect strong quantum gravity corrections. So it is possible that there will be a lower bound to the energy (for fixed bubble size) in the full quantum theory.

We have also shown that if one tries to embed one of these small bubbles into AdS, the energy remains bounded from below, consistent with the AdS/CFT correspondence. It remains an open problem to prove that the (lowest energy) static solution with $S^2\times S^1\times\mathbb{R}$ has the minimum energy among all solutions with these boundary conditions \cite{Copsey:2006br}. For the analogous case of $S^1\times\mathbb{R}^n$ boundary conditions, it was recently proven that the static AdS soliton \cite{Horowitz:1998ha} minimizes the energy among all solutions with bulk topology $\mathbb{R}^{n+1}$ (for $3\le n\le 7$) \cite{Brendle:2024}. It would be interesting to explore whether that approach can be generalized to other boundary topologies.

\begin{acknowledgments}
We thank Diandian Wang and Xiaohua Ye for discussions at an early stage of this project. The authors were supported in part by NSF Grant PHY-2408110.
\end{acknowledgments}

\appendix

\section{\label{app:MassGen}General Expression for the Energy}
With the metric functions written as in Eq.\ \eqref{eq:mab} and Eq.\ \eqref{eq:gensol}, we show that the general expression of the ADM energy is given by Eq.\ \eqref{eq:massgen}.

We expand for large $\rho$ using Eqs.\ \eqref{eq:h12} and \eqref{eq:ae}:
\begin{align}
    h_1(\rho)&=1+\mathcal{O}\left(\rho^{-3}\right),\\
    h_2(\rho)&=\rho\left(1+\frac{3\alpha_1}{4\rho}+\frac{\alpha_2}{2\rho^2}+\mathcal{O}\left(\rho^{-3}\right)\right),\\
    \frac{h_1(\rho)}{h_2(\rho)}&=\frac{1}{\rho}\left(1-\frac{3\alpha_1}{4\rho}+\frac{9\alpha_1^2-8\alpha_2}{16\rho^2}+\mathcal{O}\left(\rho^{-3}\right)\right).
\end{align}
Therefore, we have
\begin{equation}
    \int_{\rho_0}^{\rho}\frac{h_1(x)}{h_2(x)}dx=\ln\left(\frac{\rho}{\rho_0}\right)+I_1+\frac{3\alpha_1}{4\rho}+\frac{8\alpha_2-9\alpha_1^2}{32\rho^2}+\mathcal{O}\left(\rho^{-3}\right),
\end{equation}
where
\begin{equation}
    I_1=\int_{\rho_0}^{\infty}\left(\frac{h_1(x)}{h_2(x)}-\frac{1}{x}\right)dx.
\end{equation}
Thus,
\begin{align}
    \exp\left(\int_{\rho_0}^{\rho}\frac{h_1(x)}{h_2(x)}dx\right)&=e^{I_1}\frac{\rho}{\rho_0}\left(1+\frac{3\alpha_1}{4\rho}+\frac{\alpha_2}{4\rho^2}+\mathcal{O}\left(\rho^{-3}\right)\right),\\
    \frac{\exp\left(\int_{\rho_0}^{\rho}\frac{h_1(x)}{h_2(x)}dx\right)}{h_2(\rho)}&=\frac{e^{I_1}}{\rho_0}\left(1-\frac{\alpha_2}{4\rho^2}+\mathcal{O}\left(\rho^{-3}\right)\right)\label{eq:I2i},\\
    \int_{\rho_0}^{\rho}\frac{\exp\left(\int_{\rho_0}^{y}\frac{h_1(x)}{h_2(x)}dx\right)}{h_2(y)}dy&=\frac{e^{I_1}}{\rho_0}\left(\rho-\rho_0+I_2+\frac{\alpha_2}{4\rho}+\mathcal{O}\left(\rho^{-2}\right)\right),
\end{align}
where
\begin{equation}
    I_2=\int_{\rho_0}^{\infty}\left(\frac{\rho_0\exp\left(\int_{\rho_0}^{y}\frac{h_1(x)}{h_2(x)}dx\right)}{e^{I_1}h_2(y)}-1\right)dy.
\end{equation}
Therefore,
\begin{equation}
    \beta(\rho)=1+\frac{1}{\rho}\left(\frac{\rho_0c}{e^{I_1}}-\frac{3\alpha_1}{4}-\rho_0+I_2\right)+\mathcal{O}\left(\rho^{-2}\right).
\end{equation}
Eq.\ \eqref{eq:massgen} then follows from Eqs.\ \eqref{eq:f12}, \eqref{eq:mass}, and \eqref{eq:mab}.

\section{\label{app:PolA}Another Family of Initial Data}
Our example in the main text has a curvature that grows exponentially with the energy. In this appendix, we provide another example that also has arbitrarily negative energy for any fixed $\rho_0$ and $R$ but whose curvature grows only polynomially with the energy.

Consider the ansatz
\begin{equation}
    \alpha(\rho)=\frac{(\rho-\rho_0)(\rho-\rho_0+d)}{(\rho-\rho_0)^2+a(\rho-\rho_0)+b^2}\exp\left(-\frac{p}{\rho}\right),
\end{equation}
where $d>0$, $b>0$, $a>-2b$, $p\in\mathbb{R}$. We calculate
\begin{equation}
    \alpha'(\rho_0)=\frac{d}{b^2}\exp\left(-\frac{p}{\rho_0}\right),\quad\alpha_1=d-a-p,\quad\alpha_2=a(a-d)-b^2+(a-d)p+\frac{p^2}{2}-(a-d)\rho_0.
\end{equation}
We can use the partial fraction decomposition to write
\begin{equation}
    \frac{h_1(\rho)}{h_2(\rho)}=\frac{2p}{\rho^2}-\frac{4}{\rho}-\frac{4(2(\rho-\rho_0)+a)}{(\rho-\rho_0)^2+a(\rho-\rho_0)+b^2}+\frac{P(\rho-\rho_0)}{Q(\rho-\rho_0)},
\end{equation}
where
\begin{align}
    &P(x)=13x^4+(13a+9d+2p+8\rho_0)x^3+\left(12b^2+9a(d+\rho_0)+\frac{3(a+d)p}{2}+3d\rho_0\right)x^2\notag\\
    &\quad+(8b^2(d+\rho_0)+b^2p+adp+4ad\rho_0+(a-d)\rho_0^2)x+\frac{b^2dp}{2}+3b^2d\rho_0+b^2\rho_0^2,\\
    &Q(x)=x^5+\frac{5a+3d+p+4\rho_0}{4}x^4+\left(\frac{3b^2}{2}+ad+\frac{(a+d)p}{4}+\frac{3a\rho_0}{2}+\frac{d\rho_0}{2}\right)x^3\notag\\
    &\quad+\frac{(5d+p+8\rho_0)b^2+4ad\rho_0+(a-d)\rho_0^2+adp}{4}x^2+\frac{(dp+6d\rho_0+2\rho_0^2)b^2}{4}x+\frac{b^2d\rho_0^2}{4}.
\end{align}
We can further decompose the expression into
\begin{equation}
    \frac{h_1(\rho)}{h_2(\rho)}=\frac{2p}{\rho^2}-\frac{4}{\rho}-\frac{4}{\rho-\rho_0-\rho_+}-\frac{4}{\rho-\rho_0-\rho_-}+\sum_{i=1}^{5}\frac{p_i}{\rho-\rho_0-\rho_i},
\end{equation}
where $\rho_{\pm}$ are the roots of $x^2+ax+b^2$, $\rho_i$ are the roots of $Q(x)$, and $p_{i}=P(\rho_i)/Q'(\rho_i)$. Thus,
\begin{equation}
    \exp\left(\int_{\rho_0}^{\rho}\frac{h_1(x)}{h_2(x)}dx\right)=\left(\frac{\rho_0b^2}{\rho((\rho-\rho_0)^2+a(\rho-\rho_0)+b^2)}\right)^4\left(\frac{\rho-\rho_0-\rho_i}{-\rho_i}\right)^{p_i}e^{\frac{2p}{\rho_0}-\frac{2p}{\rho}}.
\end{equation}
Using the same argument as in section \ref{sec:Sol}, $\sum_{i=1}^{5}p_i=13$. Therefore,
\begin{equation}
    e^{I_1}=\rho_0^5b^8\prod_{i=1}^{5}\left(-\rho_i\right)^{-p_i}\exp\left(\frac{2p}{\rho_0}\right).
\end{equation}
Consider the scenario of large $b$. We can use Vieta's formulas to show that
\begin{subequations}
\begin{align}
    \rho_1&=r_1+\mathcal{O}\left(b^{-1/2}\right),\quad\rho_2=r_2+\mathcal{O}\left(b^{-1/2}\right),\quad\rho_3=r_3+\mathcal{O}\left(b^{-1/2}\right),\\
    \rho_4&=i\sqrt{\frac{3}{2}}b+\frac{4\rho_0+d-15a-p}{24}+\mathcal{O}\left(b^{-1/2}\right),\\
    \rho_5&=-i\sqrt{\frac{3}{2}}b+\frac{4\rho_0+d-15a-p}{24}+\mathcal{O}\left(b^{-1/2}\right),
\end{align}
\end{subequations}
where $r_i$ are the roots of the polynomial
\begin{equation}
    6x^3+(5d+p+8\rho_0)x^2+(dp+6d\rho_0+2\rho_0^2)x+d\rho_0^2.
\end{equation}
Thus,
\begin{subequations}
\begin{align}
    p_1&=q_1+\mathcal{O}\left(b^{-1/2}\right),\quad p_2=q_2+\mathcal{O}\left(b^{-1/2}\right),\quad p_3=q_3+\mathcal{O}\left(b^{-1/2}\right),\\
    p_4&=\frac{5}{2}+\mathcal{O}\left(b^{-1/2}\right),\quad p_5=\frac{5}{2}+\mathcal{O}\left(b^{-1/2}\right),
\end{align}
\end{subequations}
where $q_i$ are given by
\begin{equation}
    q_i=\frac{48r_i^2+4(8\rho_0+8d+p)r_i+4\rho_0^2+12d\rho_0+2dp}{18r_i^2+2(8\rho_0+5d+p)r_i+2\rho_0^2+6d\rho_0+dp}.
\end{equation}
Hence,
\begin{equation}
    -\frac{2\rho_0}{R^2e^{I_1}\left(\alpha'(\rho_0)\right)^2}=-\frac{3^{5/2}\prod_{i=1}^{3}\left(-r_i\right)^{q_i}b}{2^{3/2}R^2\rho_0^4d^2}\left(1+\mathcal{O}\left(b^{-1/2}\right)\right).
\end{equation}
So this contribution to the energy \eqref{eq:massgen} grows linearly in $b$.

Now let us consider the slope. For large $p$, we can again use Vieta's formulas to obtain
\begin{subequations}
\begin{align}
    r_1&=-\frac{p}{6}+\frac{d-8\rho_0}{6}+\frac{2\rho_0^2+d^2-2d\rho_0}{p}+\mathcal{O}\left(p^{-2}\right),\\
    r_2&=-d-\frac{(d-\rho_0)^2}{p}+\mathcal{O}\left(p^{-2}\right),\\
    r_3&=-\frac{\rho_0^2}{p}+\mathcal{O}\left(p^{-2}\right).
\end{align}
\end{subequations}
Thus,
\begin{equation}
    q_1=4+\mathcal{O}\left(p^{-1}\right),\quad q_2=2+\mathcal{O}\left(p^{-1}\right),\quad q_3=2+\mathcal{O}\left(p^{-1}\right).
\end{equation}
Hence, for $b\gg p\gg0$, to leading order,
\begin{equation}
    -\frac{2\rho_0}{R^2e^{I_1}\left(\alpha'(\rho_0)\right)^2}=-\frac{p^{2}b}{2^{11/2}3^{3/2}R^2}.
\end{equation}
On the other hand, by the same reasoning as in section \ref{sec:Sol}, $I_2$ is at most linear in $b$. But if it is linear, the slope is determined by the asymptotic behavior \eqref{eq:I2i} which is independent of $p$ since $\alpha_2=-b^2$ to leading order. (We have checked that higher order terms in the large $\rho$ expansion do not introduce $p$-dependence.) Therefore, for any fixed $\rho_0$ and $R$, it is always possible to choose a (fixed) $p$ large enough such that $E\sim-b$ for large $b$. Further, the curvature grows polynomially with $b$.

\bibliography{Horowitz-Lu}

\begin{thebibliography}{21}%
\makeatletter
\providecommand \@ifxundefined [1]{%
 \@ifx{#1\undefined}
}%
\providecommand \@ifnum [1]{%
 \ifnum #1\expandafter \@firstoftwo
 \else \expandafter \@secondoftwo
 \fi
}%
\providecommand \@ifx [1]{%
 \ifx #1\expandafter \@firstoftwo
 \else \expandafter \@secondoftwo
 \fi
}%
\providecommand \natexlab [1]{#1}%
\providecommand \enquote  [1]{``#1''}%
\providecommand \bibnamefont  [1]{#1}%
\providecommand \bibfnamefont [1]{#1}%
\providecommand \citenamefont [1]{#1}%
\providecommand \href@noop [0]{\@secondoftwo}%
\providecommand \href [0]{\begingroup \@sanitize@url \@href}%
\providecommand \@href[1]{\@@startlink{#1}\@@href}%
\providecommand \@@href[1]{\endgroup#1\@@endlink}%
\providecommand \@sanitize@url [0]{\catcode `\\12\catcode `\$12\catcode `\&12\catcode `\#12\catcode `\^12\catcode `\_12\catcode `\%12\relax}%
\providecommand \@@startlink[1]{}%
\providecommand \@@endlink[0]{}%
\providecommand \url  [0]{\begingroup\@sanitize@url \@url }%
\providecommand \@url [1]{\endgroup\@href {#1}{\urlprefix }}%
\providecommand \urlprefix  [0]{URL }%
\providecommand \Eprint [0]{\href }%
\providecommand \doibase [0]{https://doi.org/}%
\providecommand \selectlanguage [0]{\@gobble}%
\providecommand \bibinfo  [0]{\@secondoftwo}%
\providecommand \bibfield  [0]{\@secondoftwo}%
\providecommand \translation [1]{[#1]}%
\providecommand \BibitemOpen [0]{}%
\providecommand \bibitemStop [0]{}%
\providecommand \bibitemNoStop [0]{.\EOS\space}%
\providecommand \EOS [0]{\spacefactor3000\relax}%
\providecommand \BibitemShut  [1]{\csname bibitem#1\endcsname}%
\let\auto@bib@innerbib\@empty
%</preamble>
\bibitem [{\citenamefont {Schoen}\ and\ \citenamefont {Yau}(1979{\natexlab{a}})}]{Schoen:1979rg}%
  \BibitemOpen
  \bibfield  {author} {\bibinfo {author} {\bibfnamefont {R.}~\bibnamefont {Schoen}}\ and\ \bibinfo {author} {\bibfnamefont {S.-T.}\ \bibnamefont {Yau}},\ }\bibfield  {title} {\bibinfo {title} {{On the proof of the positive mass conjecture in general relativity}},\ }\href {https://doi.org/10.1007/BF01940959} {\bibfield  {journal} {\bibinfo  {journal} {Commun. Math. Phys.}\ }\textbf {\bibinfo {volume} {65}},\ \bibinfo {pages} {45} (\bibinfo {year} {1979}{\natexlab{a}})}\BibitemShut {NoStop}%
\bibitem [{\citenamefont {Schoen}\ and\ \citenamefont {Yau}(1979{\natexlab{b}})}]{Schoen:1979zz}%
  \BibitemOpen
  \bibfield  {author} {\bibinfo {author} {\bibfnamefont {R.}~\bibnamefont {Schoen}}\ and\ \bibinfo {author} {\bibfnamefont {S.-T.}\ \bibnamefont {Yau}},\ }\bibfield  {title} {\bibinfo {title} {{Positivity of the total mass of a general space-time}},\ }\href {https://doi.org/10.1103/PhysRevLett.43.1457} {\bibfield  {journal} {\bibinfo  {journal} {Phys. Rev. Lett.}\ }\textbf {\bibinfo {volume} {43}},\ \bibinfo {pages} {1457} (\bibinfo {year} {1979}{\natexlab{b}})}\BibitemShut {NoStop}%
\bibitem [{\citenamefont {Schoen}\ and\ \citenamefont {Yau}(1981)}]{Schoen:1981vd}%
  \BibitemOpen
  \bibfield  {author} {\bibinfo {author} {\bibfnamefont {R.}~\bibnamefont {Schoen}}\ and\ \bibinfo {author} {\bibfnamefont {S.-T.}\ \bibnamefont {Yau}},\ }\bibfield  {title} {\bibinfo {title} {{Proof of the positive mass theorem. II.}},\ }\href {https://doi.org/10.1007/BF01942062} {\bibfield  {journal} {\bibinfo  {journal} {Commun. Math. Phys.}\ }\textbf {\bibinfo {volume} {79}},\ \bibinfo {pages} {231} (\bibinfo {year} {1981})}\BibitemShut {NoStop}%
\bibitem [{\citenamefont {Witten}(1981)}]{Witten:1981mf}%
  \BibitemOpen
  \bibfield  {author} {\bibinfo {author} {\bibfnamefont {E.}~\bibnamefont {Witten}},\ }\bibfield  {title} {\bibinfo {title} {A new proof of the positive energy theorem},\ }\href {https://doi.org/10.1007/BF01208277} {\bibfield  {journal} {\bibinfo  {journal} {Commun. Math. Phys.}\ }\textbf {\bibinfo {volume} {80}},\ \bibinfo {pages} {381} (\bibinfo {year} {1981})}\BibitemShut {NoStop}%
\bibitem [{\citenamefont {Kaluza}(1921)}]{Kaluza:1921tu}%
  \BibitemOpen
  \bibfield  {author} {\bibinfo {author} {\bibfnamefont {T.}~\bibnamefont {Kaluza}},\ }\bibfield  {title} {\bibinfo {title} {{Zum Unit{\"a}tsproblem der Physik}},\ }\href {https://doi.org/10.1142/S0218271818700017} {\bibfield  {journal} {\bibinfo  {journal} {Sitzungsber. Preuss. Akad. Wiss. Berlin (Math. Phys. )}\ }\textbf {\bibinfo {volume} {1921}},\ \bibinfo {pages} {966} (\bibinfo {year} {1921})},\ \Eprint {https://arxiv.org/abs/1803.08616} {arXiv:1803.08616 [physics.hist-ph]} \BibitemShut {NoStop}%
\bibitem [{\citenamefont {Klein}(1926)}]{Klein1926}%
  \BibitemOpen
  \bibfield  {author} {\bibinfo {author} {\bibfnamefont {O.}~\bibnamefont {Klein}},\ }\bibfield  {title} {\bibinfo {title} {{Quantentheorie} und fünfdimensionale {Relativitätstheorie}},\ }\href {https://doi.org/10.1007/BF01397481} {\bibfield  {journal} {\bibinfo  {journal} {Zeitschrift für Physik A}\ }\textbf {\bibinfo {volume} {37}},\ \bibinfo {pages} {895} (\bibinfo {year} {1926})}\BibitemShut {NoStop}%
\bibitem [{\citenamefont {Zhang}(1999)}]{Zhang:1999ma}%
  \BibitemOpen
  \bibfield  {author} {\bibinfo {author} {\bibfnamefont {X.}~\bibnamefont {Zhang}},\ }\bibfield  {title} {\bibinfo {title} {{Positive mass conjecture for five-dimensional Lorentzian manifolds}},\ }\href {https://doi.org/10.1063/1.532906} {\bibfield  {journal} {\bibinfo  {journal} {J. Math. Phys.}\ }\textbf {\bibinfo {volume} {40}},\ \bibinfo {pages} {3540} (\bibinfo {year} {1999})}\BibitemShut {NoStop}%
\bibitem [{\citenamefont {Dai}(2004)}]{Dai_2004}%
  \BibitemOpen
  \bibfield  {author} {\bibinfo {author} {\bibfnamefont {X.}~\bibnamefont {Dai}},\ }\bibfield  {title} {\bibinfo {title} {A positive mass theorem for spaces with asymptotic {SUSY} compactification},\ }\href {https://doi.org/10.1007/s00220-003-0986-2} {\bibfield  {journal} {\bibinfo  {journal} {Communications in Mathematical Physics}\ }\textbf {\bibinfo {volume} {244}},\ \bibinfo {pages} {335–345} (\bibinfo {year} {2004})}\BibitemShut {NoStop}%
\bibitem [{\citenamefont {Dai}(2005)}]{Dai_2005}%
  \BibitemOpen
  \bibfield  {author} {\bibinfo {author} {\bibfnamefont {X.}~\bibnamefont {Dai}},\ }\bibfield  {title} {\bibinfo {title} {A note on positive energy theorem for spaces with asymptotic {SUSY} compactification},\ }\bibfield  {journal} {\bibinfo  {journal} {Journal of Mathematical Physics}\ }\textbf {\bibinfo {volume} {46}},\ \href {https://doi.org/10.1063/1.1862095} {10.1063/1.1862095} (\bibinfo {year} {2005})\BibitemShut {NoStop}%
\bibitem [{\citenamefont {Witten}(1982)}]{Witten:1981gj}%
  \BibitemOpen
  \bibfield  {author} {\bibinfo {author} {\bibfnamefont {E.}~\bibnamefont {Witten}},\ }\bibfield  {title} {\bibinfo {title} {{Instability of the Kaluza-Klein vacuum}},\ }\href {https://doi.org/10.1016/0550-3213(82)90007-4} {\bibfield  {journal} {\bibinfo  {journal} {Nucl. Phys. B}\ }\textbf {\bibinfo {volume} {195}},\ \bibinfo {pages} {481} (\bibinfo {year} {1982})}\BibitemShut {NoStop}%
\bibitem [{\citenamefont {Brill}\ and\ \citenamefont {Horowitz}(1991)}]{Brill:1991qe}%
  \BibitemOpen
  \bibfield  {author} {\bibinfo {author} {\bibfnamefont {D.}~\bibnamefont {Brill}}\ and\ \bibinfo {author} {\bibfnamefont {G.~T.}\ \bibnamefont {Horowitz}},\ }\bibfield  {title} {\bibinfo {title} {{Negative energy in string theory}},\ }\href {https://doi.org/10.1016/0370-2693(91)90618-Z} {\bibfield  {journal} {\bibinfo  {journal} {Phys. Lett. B}\ }\textbf {\bibinfo {volume} {262}},\ \bibinfo {pages} {437} (\bibinfo {year} {1991})}\BibitemShut {NoStop}%
\bibitem [{\citenamefont {Brill}\ and\ \citenamefont {Pfister}(1989)}]{Brill:1989di}%
  \BibitemOpen
  \bibfield  {author} {\bibinfo {author} {\bibfnamefont {D.}~\bibnamefont {Brill}}\ and\ \bibinfo {author} {\bibfnamefont {H.}~\bibnamefont {Pfister}},\ }\bibfield  {title} {\bibinfo {title} {{States of negative total energy in Kaluza-Klein theory}},\ }\href {https://doi.org/10.1016/0370-2693(89)91559-1} {\bibfield  {journal} {\bibinfo  {journal} {Phys. Lett. B}\ }\textbf {\bibinfo {volume} {228}},\ \bibinfo {pages} {359} (\bibinfo {year} {1989})}\BibitemShut {NoStop}%
\bibitem [{\citenamefont {Bombelli}\ \emph {et~al.}(1987)\citenamefont {Bombelli}, \citenamefont {Koul}, \citenamefont {Kunstatter}, \citenamefont {Lee},\ and\ \citenamefont {Sorkin}}]{Bombelli:1986sb}%
  \BibitemOpen
  \bibfield  {author} {\bibinfo {author} {\bibfnamefont {L.}~\bibnamefont {Bombelli}}, \bibinfo {author} {\bibfnamefont {R.~K.}\ \bibnamefont {Koul}}, \bibinfo {author} {\bibfnamefont {G.}~\bibnamefont {Kunstatter}}, \bibinfo {author} {\bibfnamefont {J.}~\bibnamefont {Lee}},\ and\ \bibinfo {author} {\bibfnamefont {R.~D.}\ \bibnamefont {Sorkin}},\ }\bibfield  {title} {\bibinfo {title} {On energy in five-dimensional gravity and the mass of the {Kaluza-Klein} monopole},\ }\href {https://doi.org/10.1016/0550-3213(87)90404-4} {\bibfield  {journal} {\bibinfo  {journal} {Nucl. Phys. B}\ }\textbf {\bibinfo {volume} {289}},\ \bibinfo {pages} {735} (\bibinfo {year} {1987})}\BibitemShut {NoStop}%
\bibitem [{\citenamefont {Deser}\ and\ \citenamefont {Soldate}(1989)}]{Deser:1988fc}%
  \BibitemOpen
  \bibfield  {author} {\bibinfo {author} {\bibfnamefont {S.}~\bibnamefont {Deser}}\ and\ \bibinfo {author} {\bibfnamefont {M.}~\bibnamefont {Soldate}},\ }\bibfield  {title} {\bibinfo {title} {Gravitational energy in spaces with compactified dimensions},\ }\href {https://doi.org/10.1016/0550-3213(89)90175-2} {\bibfield  {journal} {\bibinfo  {journal} {Nucl. Phys. B}\ }\textbf {\bibinfo {volume} {311}},\ \bibinfo {pages} {739} (\bibinfo {year} {1989})}\BibitemShut {NoStop}%
\bibitem [{\citenamefont {Corley}\ and\ \citenamefont {Jacobson}(1994)}]{Corley:1994mc}%
  \BibitemOpen
  \bibfield  {author} {\bibinfo {author} {\bibfnamefont {S.}~\bibnamefont {Corley}}\ and\ \bibinfo {author} {\bibfnamefont {T.}~\bibnamefont {Jacobson}},\ }\bibfield  {title} {\bibinfo {title} {{Collapse of Kaluza-Klein bubbles}},\ }\href {https://doi.org/10.1103/PhysRevD.49.R6261} {\bibfield  {journal} {\bibinfo  {journal} {Phys. Rev. D}\ }\textbf {\bibinfo {volume} {49}},\ \bibinfo {pages} {R6261} (\bibinfo {year} {1994})},\ \Eprint {https://arxiv.org/abs/gr-qc/9403017} {arXiv:gr-qc/9403017} \BibitemShut {NoStop}%
\bibitem [{\citenamefont {Copsey}\ and\ \citenamefont {Horowitz}(2006)}]{Copsey:2006br}%
  \BibitemOpen
  \bibfield  {author} {\bibinfo {author} {\bibfnamefont {K.}~\bibnamefont {Copsey}}\ and\ \bibinfo {author} {\bibfnamefont {G.~T.}\ \bibnamefont {Horowitz}},\ }\bibfield  {title} {\bibinfo {title} {{Gravity dual of gauge theory on $S^2\times S^1\times R$}},\ }\href {https://doi.org/10.1088/1126-6708/2006/06/021} {\bibfield  {journal} {\bibinfo  {journal} {JHEP}\ }\textbf {\bibinfo {volume} {06}},\ \bibinfo {pages} {021}},\ \Eprint {https://arxiv.org/abs/hep-th/0602003} {arXiv:hep-th/0602003} \BibitemShut {NoStop}%
\bibitem [{\citenamefont {Horowitz}\ \emph {et~al.}(2024)\citenamefont {Horowitz}, \citenamefont {Wang},\ and\ \citenamefont {Ye}}]{Horowitz:2024hqq}%
  \BibitemOpen
  \bibfield  {author} {\bibinfo {author} {\bibfnamefont {G.~T.}\ \bibnamefont {Horowitz}}, \bibinfo {author} {\bibfnamefont {D.}~\bibnamefont {Wang}},\ and\ \bibinfo {author} {\bibfnamefont {X.}~\bibnamefont {Ye}},\ }\bibfield  {title} {\bibinfo {title} {{New energy inequality in AdS spacetimes}},\ }\href {https://doi.org/10.1103/PhysRevD.110.064015} {\bibfield  {journal} {\bibinfo  {journal} {Phys. Rev. D}\ }\textbf {\bibinfo {volume} {110}},\ \bibinfo {pages} {064015} (\bibinfo {year} {2024})},\ \Eprint {https://arxiv.org/abs/2406.13068} {arXiv:2406.13068 [gr-qc]} \BibitemShut {NoStop}%
\bibitem [{\citenamefont {Balasubramanian}\ and\ \citenamefont {Kraus}(1999)}]{Balasubramanian:1999re}%
  \BibitemOpen
  \bibfield  {author} {\bibinfo {author} {\bibfnamefont {V.}~\bibnamefont {Balasubramanian}}\ and\ \bibinfo {author} {\bibfnamefont {P.}~\bibnamefont {Kraus}},\ }\bibfield  {title} {\bibinfo {title} {{A stress tensor for anti-de Sitter gravity}},\ }\href {https://doi.org/10.1007/s002200050764} {\bibfield  {journal} {\bibinfo  {journal} {Commun. Math. Phys.}\ }\textbf {\bibinfo {volume} {208}},\ \bibinfo {pages} {413} (\bibinfo {year} {1999})},\ \Eprint {https://arxiv.org/abs/hep-th/9902121} {arXiv:hep-th/9902121} \BibitemShut {NoStop}%
\bibitem [{\citenamefont {Hawking}\ and\ \citenamefont {Horowitz}(1996)}]{Hawking:1995fd}%
  \BibitemOpen
  \bibfield  {author} {\bibinfo {author} {\bibfnamefont {S.~W.}\ \bibnamefont {Hawking}}\ and\ \bibinfo {author} {\bibfnamefont {G.~T.}\ \bibnamefont {Horowitz}},\ }\bibfield  {title} {\bibinfo {title} {{The gravitational Hamiltonian, action, entropy and surface terms}},\ }\href {https://doi.org/10.1088/0264-9381/13/6/017} {\bibfield  {journal} {\bibinfo  {journal} {Class. Quant. Grav.}\ }\textbf {\bibinfo {volume} {13}},\ \bibinfo {pages} {1487} (\bibinfo {year} {1996})},\ \Eprint {https://arxiv.org/abs/gr-qc/9501014} {arXiv:gr-qc/9501014} \BibitemShut {NoStop}%
\bibitem [{\citenamefont {Horowitz}\ and\ \citenamefont {Myers}(1998)}]{Horowitz:1998ha}%
  \BibitemOpen
  \bibfield  {author} {\bibinfo {author} {\bibfnamefont {G.~T.}\ \bibnamefont {Horowitz}}\ and\ \bibinfo {author} {\bibfnamefont {R.~C.}\ \bibnamefont {Myers}},\ }\bibfield  {title} {\bibinfo {title} {{The AdS / CFT correspondence and a new positive energy conjecture for general relativity}},\ }\href {https://doi.org/10.1103/PhysRevD.59.026005} {\bibfield  {journal} {\bibinfo  {journal} {Phys. Rev. D}\ }\textbf {\bibinfo {volume} {59}},\ \bibinfo {pages} {026005} (\bibinfo {year} {1998})},\ \Eprint {https://arxiv.org/abs/hep-th/9808079} {arXiv:hep-th/9808079} \BibitemShut {NoStop}%
\bibitem [{\citenamefont {Brendle}\ and\ \citenamefont {Hung}(2024)}]{Brendle:2024}%
  \BibitemOpen
  \bibfield  {author} {\bibinfo {author} {\bibfnamefont {S.}~\bibnamefont {Brendle}}\ and\ \bibinfo {author} {\bibfnamefont {P.-K.}\ \bibnamefont {Hung}},\ }\href@noop {} {\bibinfo {title} {{Systolic inequalities and the Horowitz-Myers conjecture}}} (\bibinfo {year} {2024}),\ \Eprint {https://arxiv.org/abs/2406.04283} {arXiv:2406.04283 [math]} \BibitemShut {NoStop}%
\end{thebibliography}%

\end{document}